# Diversity-Multiplexing Tradeoff for the Multiple-Antenna Wire-tap Channel

Melda Yuksel, *Member, IEEE,* and Elza Erkip, *Fellow, IEEE*

*Abstract*— In this paper the fading multiple antenna (MIMO) wire-tap channel is investigated under short term power constraints. The *secret* diversity gain and the *secret* multiplexing gain are defined. Using these definitions, the *secret* diversity-multiplexing tradeoff (DMT) is calculated analytically for no transmitter side channel state information (CSI) and for full CSI. When there is no CSI at the transmitter, under the assumption of Gaussian codebooks, it is shown that the eavesdropper *steals* both transmitter and receiver antennas, and the secret DMT depends on the remaining degrees of freedom. When CSI is available at the transmitter (CSIT), the eavesdropper *steals* only transmitter antennas. This dependence on the availability of CSI is unlike the DMT results without secrecy constraints, where the DMT remains the same for no CSI and full CSI at the transmitter under short term power constraints. A *zero-forcing* type scheme is shown to achieve the secret DMT when CSIT is available.

**Keywords: Diversity-multiplexing tradeoff, MIMO, secrecy, wire-tap channel.**

## I. INTRODUCTION

In wireless communications, communication medium is shared. Any transmission can be overheard by nearby nodes. If eavesdroppers are present in the environment, then all confidential information such as user IDs, passwords, or credit card numbers become vulnerable. In addition to voice, image, video, and data transmissions, future applications envision wireless transmission of sensitive information such as personal and locality information. Therefore, wireless security is an essential system requirement.

In current wireless systems, protection against eavesdropping is provided at higher layers of the Open Systems Interconnection (OSI) reference model. Transport, network or application layer protocols aim to prevent eavesdropping using encryption. However, key exchange and renewal may be difficult considering the wireless network dynamics. Thus, developing new security protocols, which do not necessitate keys or cannot be broken even with infinite computing power, are of utmost importance for future development of wireless applications.

Physical layer security techniques provide unconditional secrecy through channel coding at the physical layer and complement higher layer security methods. The wire-tap channel, which is one of the building blocks of information-theoretic security, was introduced in [1], and later studied in [2] and [3]. A long gap of about 30 years followed these initial papers until the research community regained interest in secure communication applications for wireless networks. Recently, the ergodic secrecy capacity is calculated for fading wire-tap channels in [4], [5]. The multiple access channel with an external eavesdropper is studied in [6] and [7]. Secrecy capacity for broadcast and interference channels is investigated in [8], and for fading broadcast channels in [9]. Similarly, the secrecy capacity for the relay channel with an external eavesdropper is studied in [10] and [11].

In wireless channels, multiple antennas increase robustness against fading, and also transmission rates. Multiple antennas are considered in the context of wire-tap channels in [12], [13]-[17]. In [13] the authors find the secrecy capacity of the Gaussian multiple-input multiple-output (MIMO) wire-tap channel, when the source and the destination have two antennas each and the eavesdropper has only a single antenna. Concurrent work in [14] and [15] establish the secrecy capacity for the fading MIMO wire-tap channel under the full CSI assumption for arbitrary antenna numbers. A closed-form expression for the secrecy capacity is found in [18].

For fading channels under stringent delay constraints, the outage formulation proves to be more useful than capacity. For the wire-tap channel, outage approach is considered in [12], [19] and [20]. Outage probability for a target secrecy rate is also investigated in [5], when the source, the destination and the eavesdropper have CSI, and optimal power allocation policies that minimize the outage probability are calculated. On the other hand, the notion of *secure degrees of freedom* are investigated in [21], [22], [23], [24], [25], [26] and [27].

An important performance measure for MIMO fading channels that simultaneously considers probability of error and data rates is the diversity-multiplexing tradeoff (DMT), established in [28]. The DMT is a high SNR analysis and describes the fundamental tradeoff between the diversity gain and the multiplexing gain. The diversity gain is the decay rate of the probability of error, and the multiplexing gain is the rate of increase of the transmission rate in the limit of high SNR. The DMT is strongly related to the probability of outage as probability of error is generally dominated by the outage event at high SNR.

In this paper we investigate the multiple-antenna wire-tap channel from the DMT perspective. We define the *secret* multiplexing gain, the *secret* diversity gain and the *secret* DMT. We argue that the eavesdropper can be thought of as "stealing" degrees of freedom from the source-destination channel, and the *secret* DMT depends on the remaining degrees of freedom, when there is no CSIT. This behavior is also observed in [21] for compound channels only for the maximum multiplexing gain point. Our work can be thought of as a generalization of [21], capturing the behavior for all diversity gains. We also argue that the secret DMT depends on the available CSI at the transmitter (CSIT). This is unlike the regular point-to-point DMT without security constraints, which is not affected from the transmitter CSI for constant-rate transmission. Under CSIT assumptions, we also suggest a *zero-forcing* type scheme, which achieves the secret DMT upper bounds.

Next, we introduce the system model in Section II and then state the secret DMT for no CSIT in Section III. Section IV covers the secret DMT when there is CSIT. We conclude in Section V.

## II. System Model and Preliminaries

We consider a multiple-antenna wire-tap channel, in which the source, the destination and the eavesdropper have $m$, $n$ and $k$ antennas respectively. Both the destination and the eavesdropper have CSI about their incoming channels. In Section III we assume the source node does not have any transmit CSI. We will consider the case when the source has transmitter CSI in Section IV.

For each channel use the channel is represented as follows:

$$\mathbf{Y}_D = \mathbf{H}_D \mathbf{X} + \mathbf{Z}_D \qquad (1)$$
$$\mathbf{Y}_E = \mathbf{H}_E \mathbf{X} + \mathbf{Z}_E. \qquad (2)$$

In the above equations $\mathbf{X}$ is an $m \times 1$ vector, which denotes the transmitted source signal. $\mathbf{Y}_D$ and $\mathbf{Y}_E$ are $n \times 1$ and $k \times 1$ vectors, and represent the received signals at the destination and the eavesdropper respectively. Similarly, $\mathbf{Z}_D$ and $\mathbf{Z}_E$ are $n \times 1$, and $k \times 1$ vectors that indicate the independent additive noise at the destination and the eavesdropper. Both $\mathbf{Z}_D$ and $\mathbf{Z}_E$ have independent and identically distributed (i.i.d.) complex Gaussian entries with zero mean and unit variance. The matrices $\mathbf{H}_D$ and $\mathbf{H}_E$, consisting of i.i.d. complex Gaussian entries with zero mean and unit variance, are of size $n \times m$, and $k \times m$. They respectively denote the channel gains between the source and the destination and the source and the eavesdropper. As the fading is assumed to be slow, $\mathbf{H}_D$ and $\mathbf{H}_E$ are fixed for the whole duration of the communication.

When there is no secrecy constraint, the source fixes its transmission rate at $R^{(T)}(\text{SNR})$ and aims to transmit the message $W$, $W \in \mathcal{W} = \{1, 2, ..., 2^{NR^{(T)}}\}$, in $N$ channel uses. The destination declares an error if its decision $\hat{W} \neq W$, $\hat{W} \in \mathcal{W}$. This error probability, $P_e(\text{SNR})$, is shown to be dominated by the outage event, or $P_e(\text{SNR}) \doteq P(R < R^{(T)}(\text{SNR}))^1$, where $R = I(\mathbf{X}; \mathbf{Y}_D)$ is the instantaneous mutual information corresponding to the chosen transmission scheme [28].

The diversity-multiplexing tradeoff, $d(r)$, establishes a relation between the target transmission rate $R^{(T)}(\text{SNR})$ and probability of error $P_e(\text{SNR})$ [28], where $r$ is the multiplexing gain. It is shown to be the piecewise linear function joining the points $d_{m,n}(l) = (m - l)(n - l)$, $l = 0, 1, ..., \min\{m, n\}$ [28]. The degrees of freedom in this system is $\min\{m, n\}$, and the multiplexing gain $r$ can increase up to this value. Similarly, the maximum diversity gain is $mn$, and the diversity gain decreases as the multiplexing gain increases.

Under secrecy constraints, the source not only aims to send the message $W$ reliably but also securely to the destination. The secrecy rate, $R_s$, is achieved if the secrecy constraint is satisfied; i.e. $R_s = \lim_{N \to \infty} \frac{1}{N} H(W) = \lim_{N \to \infty} \frac{1}{N} H(W | Y_E^N)$, and the probability of decoding error at the destination approaches zero as $N$ approaches infinity [1]. The term $\lim_{N \to \infty} \frac{1}{N} H(W | Y_E^N)$ is also known as the *equivocation rate.*

The papers [1], [2], [3] prove that the secrecy rate

$$R_s = [I(\mathbf{X}; \mathbf{Y}_D) - I(\mathbf{X}; \mathbf{Y}_E)]^+ \qquad (3)$$

is achievable for any input distribution $p(\mathbf{X})$, where $x^+$ denotes $\max\{0, x\}$. A brief overview of the achievability is as follows: Define $A = 2^{NR^{(s)}}$, $B = 2^{NR^{(d)}}$ and the sets $\mathcal{A} = \{1, ..., A\}$ and $\mathcal{B} = \{1, ..., B\}$. The source generates $A \times B$ channel codewords $\mathbf{X}_1^N$ i.i.d. with $p(\mathbf{X})$. In order to send a secret message $a \in \mathcal{A}$, the source chooses $b$ uniformly from the set $\mathcal{B}$, forms $W = (a, b)$ and maps $W$ into the channel codeword $\mathbf{X}_1^n$. Note that $B$ is the number of dummy codewords used to confuse the eavesdropper for each $a \in \mathcal{A}$. In [1], [2], [3], with full CSI at the transmitter and the receivers, $R^{(s)}$ is set to $R_s$ defined in (3), and $R^{(d)} = I(\mathbf{X}; \mathbf{Y}_E)$. Under this setup, the total number of codewords in the source codebook is $A \times B = 2^{NI(\mathbf{X}; \mathbf{Y}_D)}$, and the destination can reliably decode $W$ and hence $a$. However, the eavesdropper can only decode the index $b$ and has no information about the secret message $a$. Thus secrecy is achieved.

In this work, we investigate the high SNR behavior of the probability of error (including the probability that secrecy is not achieved) with a target secrecy rate equal to $R_s^{(T)}(\text{SNR})$. We assume the system is delay-limited and requires constant secrecy rate transmission. There is also short-term average power constraint $m\text{SNR}$ that the transmitter has to satisfy for each codeword transmitted. We define the *secret* multiplexing gain as

$$\lim_{\text{SNR} \to \infty} \frac{R_s^{(T)}(\text{SNR})}{\log \text{SNR}} \triangleq r_s.$$

The secret multiplexing gain $r_s$ shows how fast the target secrecy rate scales with increasing SNR. The *secret* diversity gain, $d_s$, is equal to

$$\lim_{\text{SNR} \to \infty} \frac{\log P_e(\text{SNR})}{\log \text{SNR}} \triangleq -d_s,$$

where $P_e(\text{SNR})$ denotes the probability of error under secrecy constraints. In this paper, we establish the tradeoff between secret diversity gain $d_s$ and the secret multiplexing gain $r_s$, $d_s(r_s)$.

In a system with secrecy constraints, the probability of error, $P_e(\text{SNR})$, is due to two events: Either the destination does not receive the secret message reliably, or secrecy is not achieved

---

[1] The expression $f_1(\text{SNR}) \doteq f_2(\text{SNR})$ is defined as $\lim_{\text{SNR} \to \infty} \log f_1(\text{SNR}) / \log \text{SNR} = \lim_{\text{SNR} \to \infty} \log f_2(\text{SNR}) / \log \text{SNR}$. In the rest of the paper, inequalities are also defined similarly.

[29]. Then

$$
\begin{aligned}
P_e(\text{SNR}) &= P\,(\text{secrecy not achieved} \\
&\qquad \text{or main channel decoding error}) \quad (4)\\
&\leq P\,(\text{secrecy not achieved}) \\
&\qquad + :\, P\,(\text{main channel decoding error}), \quad (5)
\end{aligned}
$$

where

$$
\begin{aligned}
&P(\text{secrecy not achieved}) \\
&\quad \triangleq\ P\left(\lim_{N\to\infty}\frac{1}{N}H(W|Y_E^N) < R_s^{(T)}(\text{SNR})\right). \quad (6)
\end{aligned}
$$

For the achievability scheme described above, setting $B = 2^{NR^{(d)}(\text{SNR})}$ and $A \times B = 2^{NR^{(T)}(\text{SNR})} = 2^{NR_s^{(T)}(\text{SNR}) + NR^{(d)}(\text{SNR})}$, $P(\text{secrecy not achieved})$ defined in (6) can be calculated as [1]

$$
\begin{aligned}
&P(\text{secrecy not achieved}) \\
&\quad = P(R^{(T)}(\text{SNR}) - I(\mathbf{X};\mathbf{Y}_E) < R_s^{(T)}(\text{SNR})) \\
&\quad = P\left(I(\mathbf{X};\mathbf{Y}_E) > R^{(d)}(\text{SNR})\right). \quad (7)
\end{aligned}
$$

Finally, as the main channel outage event dominates the main channel decoding error when the channel block length-$N$ is long enough and good codes are used [28],

$$
\begin{aligned}
&P(\text{main channel decoding error}) \\
&\quad \doteq\ P\,(\text{main channel outage}) \\
&\quad = P(I(\mathbf{X};\mathbf{Y}_D) < R^{(T)}(\text{SNR})). \quad (8)
\end{aligned}
$$

Note that in (7) and (8), the terms $I(\mathbf{X};\mathbf{Y}_D)$ and $I(\mathbf{X};\mathbf{Y}_E)$ are evaluated for the chosen transmission scheme determined by the codebook distribution $p(\mathbf{X})$. On the other hand, $P_e(\text{SNR})$ in (4) can be lower bounded by

$$
\begin{aligned}
P_e(\text{SNR}) &\geq P\,(\text{secrecy not achieved}) \\
&\geq P([I(\mathbf{X};\mathbf{Y}_D) - I(\mathbf{X};\mathbf{Y}_E)]^+ < R_s^{(T)}(\text{SNR})) \\
&\geq P(I(\mathbf{X};\mathbf{Y}_D) - I(\mathbf{X};\mathbf{Y}_E) < R_s^{(T)}(\text{SNR})) \\
&\triangleq P\,(\text{secrecy rate outage}), \quad (9)
\end{aligned}
$$

for the chosen achievable scheme with $p(\mathbf{X})$.

In the following we will calculate both (7) and (8) to obtain the upper bound in (5), and (9) to establish a lower bound on $P_e(\text{SNR})$. Comparing the bounds, we will establish the secret DMT. In Section III, we assume $p(\mathbf{X})$ is an isotropic Gaussian input, whereas in Section IV-A, we calculate the bounds for the best $p(\mathbf{X})$ which attains the secrecy capacity [14].

In this paper, we assume a single transmission block of $N$ channel uses under short-term power constraint. This is unlike the scenario in [20]. In [20] there are many blocks to communicate and there is long-term power constraint. The first communication block is merely used to generate a secret key, and in the next block this key is used to enhance secrecy, while another key is generated to be used in the following block. In other words, the key generation process [20] is delay-insensitive and keys generated this way are used to protect the delay-sensitive secret messages. In our system, communication session lasts a single code block, during which secrecy has to be maintained. The transmitter and the receiver have to start secure communication immediately at the beginning of the transmission block and there is not enough time to generate a secret key.

## III. No Channel State Information at the Source

When the source node does not have CSI either about its link to the destination or to the eavesdropper in the MIMO wire-tap channel, the secrecy capacity is not known. However, motivated by the fact that when all nodes in the system have complete CSI, Gaussian codebooks are optimum, [14], [15], we assume Gaussian codebooks. We also conjecture that sending independent signals at equal power at each antenna is optimal at high SNR, as all the entries of $\mathbf{H}_D$ and $\mathbf{H}_E$ respectively are identically distributed. Without CSIT, the source has no preference over one *direction* over the other for its transmission. Thus, we assume the input covariance matrix $Q$ is a diagonal matrix $Q = \text{SNR}\mathbf{I}_m$, where $\mathbf{I}_m$ indicates an identity matrix of size $m$. Then, the achievable secrecy rate in (3) becomes

$$
\begin{aligned}
R_s &= \left[\log\left|\mathbf{I}_n + \mathbf{H}_D Q \mathbf{H}_D^\dagger\right| - \log\left|\mathbf{I}_k + \mathbf{H}_E Q \mathbf{H}_E^\dagger\right|\right]^+ \\
&= \left[\log\frac{\prod_{i=1}^{L}(1+\lambda_i \text{SNR})}{\prod_{i=1}^{k}(1+\mu_i \text{SNR})}\right]^+, \quad (10)
\end{aligned}
$$

where $L = \min\{m,n\}$, $0 \leq \lambda_1 \leq \ldots \leq \lambda_L$ are the ordered eigenvalues of the matrix $\mathbf{H}_D\mathbf{H}_D^\dagger$, $0 \leq \mu_1 \leq \ldots \leq \mu_k$ are the ordered eigenvalues of the matrix $\mathbf{H}_E\mathbf{H}_E^\dagger$, and $\dagger$ denotes the conjugate transpose.

*Theorem 1:* For the multiple-antenna wire-tap channel defined in (1) and (2), with full CSI at the destination and the eavesdropper about their incoming channel gains and no CSI at the source, if $k < \min\{m,n\}$, the secret diversity-multiplexing tradeoff achieved by isotropic Gaussian codebook is a piecewise linear function joining the points $(l, d_s(l))$, where $l = 0, 1, \ldots, \min\{m,n\} - k$ and

$$
d_s(l) = (m-k-l)(n-k-l).
$$

If $k \geq \min\{m,n\}$, then the secret diversity-multiplexing tradeoff reduces to the single point $(0,0)$.

*Proof:* We first find an upper bound on secret DMT. To do this, we calculate the probability of the secrecy rate outage of (9) for $R_s^{(T)} = r_s \log \text{SNR}$, and show that at high SNR, this probability is on the order of $\text{SNR}^{-d_{m-k,n-k}(r_s)}$. The details of the computation are presented in Appendix I.

To show that the above secret DMT upper bound is achievable, we set $R^{(T)} = R_s^{(T)} + \min\{m,k\}\log\text{SNR} = (r_s + \min\{m,k\})\log\text{SNR}$ bits/channel use, where the target secret communication rate is $R_s^{(T)}$ bits/channel use with multiplexing gain $r_s$, and $R^{(d)}(\text{SNR}) = \min\{m,k\}\log\text{SNR}$. Then the main channel is in outage when the destination cannot decode rate $R^{(T)}$, which has the probability

$$
\begin{aligned}
&P(\text{main channel outage}) \\
&\quad = P\left(I(\mathbf{X};\mathbf{Y}_D) < R^{(T)}\right) \\
&\quad = P\left(I(\mathbf{X};\mathbf{Y}_D) < (r_s + \min\{m,k\})\log\text{SNR}\right) \\
&\quad \overset{(a)}{\doteq} \text{SNR}^{-d_{m,n}(r_s + \min\{m,k\})}
\end{aligned}
$$

$$= \begin{cases} \mathrm{SNR}^{-d_{m-k,n-k}(r_s)} & \text{if } k < m \\ 1 & \text{if } k \geq m \end{cases},$$

where the mutual information is evaluated for isotropic Gaussian inputs and $(a)$ is due to [28]. On the other hand, (7) becomes

$$\begin{aligned} &P(\text{secrecy not achieved}) \\ &= P(I(\mathbf{X}; \mathbf{Y}_E) > R^{(d)}(\mathrm{SNR})) \\ &\doteq P(\min\{m, k\} \log \mathrm{SNR} < I(\mathbf{X}; \mathbf{Y}_E)) \\ &\doteq 0, \end{aligned} \quad (11)$$

since the maximum degrees of freedom in the source-eavesdropper channel is equal to $\min\{m, k\}$, [28]. Overall, if $k < m$ the upper bound on the probability of error (5) becomes

$$\begin{aligned} P_e(\mathrm{SNR}) &\doteq 0 + \mathrm{SNR}^{-d_{m-k,n-k}(r_s)} \\ &\doteq \mathrm{SNR}^{-d_{m-k,n-k}(r_s)}. \end{aligned}$$

As the lower bound on probability of error (9) is the same, we conclude that the secret DMT is equal to $d_{m-k,n-k}(r_s)$ if $k < m$. If $k \geq m$, the secret DMT is the single point $(0, 0)$. ∎

Theorem 1 states that the eavesdropper costs the system $\min\{m, k\}$ degrees of freedom, which affects the whole secret DMT curve. When the degrees of freedom in the source-eavesdropper channel, $\min\{m, k\}$, is equal to $k$, then the secret system becomes equivalent to an $(m-k) \times (n-k)$ system. However, if $\min\{m, k\} = m$, then no degrees of freedom are left for the main channel, as $m \geq \min\{m, n\}$, and the secret DMT reduces to the single point $(0, 0)$.

## IV. Channel State Information at the Source

In the previous section secret DMT is established for MIMO wire-tap channels without CSIT. In this section we assume that transmitter has perfect CSI about the channel between itself and the eavesdropper, as well as its channel to the destination. While it may be possible for the source to obtain eavesdropper CSI if both the destination and the eavesdropper are part of the same network, the full CSIT assumption may be harder to justify if the eavesdropper is merely an illegitimate listener. Nevertheless, this assumption will help us understand the limitations and properties of secret DMT. Note that secret DMT is still a meaningful metric as we consider constant secret rate applications that operate under short-term power constraints, which can suffer from outage despite the available CSIT.

In the next subsection we establish the secret DMT with CSIT and in Section IV-B we investigate different schemes that achieve the best secret DMT with CSIT.

### A. Secret DMT with CSIT

The secrecy capacity for the non-fading MIMO wire-tap channel with channel knowledge at all the terminals is found in [14], [15] as

$$C_s = \max_{\substack{Q \succeq 0, \\ \mathrm{Tr}(Q) \leq m\mathrm{SNR}}} \log \frac{\left| \mathbf{I}_n + \mathbf{H}_D Q \mathbf{H}_D^\dagger \right|}{\left| \mathbf{I}_k + \mathbf{H}_E Q \mathbf{H}_E^\dagger \right|}. \quad (12)$$

To establish the secret DMT with CSIT, we first need the following lemma.

Lemma 1: If $k < \min\{m, n\}$, then $p = \dim\{\text{Null}(\mathbf{H}_D)^\perp \cap \text{Null}(\mathbf{H}_E)\} > 0$, where $\text{Null}(\mathbf{H}_D)^\perp$ is the orthogonal complement of the null space of $\mathbf{H}_D$ and $\text{Null}(\mathbf{H}_E)$ is the null space of $\mathbf{H}_E$. If $n \leq k < m$ or $k \geq m$, then $p = 0$.

Proof: The subspaces $\text{Null}(\mathbf{H}_E)$ and $\text{Null}(\mathbf{H}_D)^\perp$ are defined in the vector space $\Re^m$. If $k < \min\{m, n\}$, then $\text{Null}(\mathbf{H}_E)$ and $\text{Null}(\mathbf{H}_D)^\perp$ respectively have dimensions $m - k$ and $q = \min\{m, n\}$, and have the basis sets $\mathcal{U} = \{\mathbf{u}_1, \mathbf{u}_2, ..., \mathbf{u}_{m-k}\}$ and $\mathcal{W} = \{\mathbf{w}_1, \mathbf{w}_2, ..., \mathbf{w}_q\}$. In other words, the sets $\mathcal{U}$ and $\mathcal{W}$ are both linearly independent sets. However, as $m - k + q > m$, $\mathcal{U} \cup \mathcal{W}$ is linearly dependent. The intersection of the hyper-planes $\mathcal{U}$ span and $\mathcal{W}$ span, includes at least one non-zero vector. Thus $p > 0$.

If $n \leq k < m$, then the basis sets $\mathcal{U}$ and $\mathcal{W}$ are same as above with $q = n$. However, in this case $\mathcal{U} \cup \mathcal{W}$ is a linearly independent set, as $m - k + q = m - k + n \leq m$. The intersection of the hyper-planes $\mathcal{U}$ span and $\mathcal{W}$ span only include $\{\mathbf{0}\}$ and thus $p = 0$.

If $k \geq m$, then $\text{Null}(\mathbf{H}_E)$ consists of only $\{\mathbf{0}\}$. Then $\text{Null}(\mathbf{H}_D)^\perp \cap \text{Null}(\mathbf{H}_E) = \{\mathbf{0}\}$, and thus $p = 0$. ∎

Theorem 2: For the multiple-antenna wire-tap channel defined in (1) and (2), with full CSI at all the terminals, if $k < m$, the secret diversity-multiplexing tradeoff, $\hat{d}_s(r_s)$ is a piecewise linear function joining the points $(l, \hat{d}_s(l))$, where $l = 0, 1, ..., m - k$ and

$$\hat{d}_s(l) = (m - k - l)(n - l).$$

If $k \geq m$, then the secret diversity-multiplexing tradeoff reduces to the single point $(0, 0)$.

Proof: When the secrecy capacity is expressed as in (12), it is hard to calculate the secret DMT. We make use of the high SNR secrecy capacity approximations provided in [14] to find the secret DMT. We investigate the three cases $k < \min\{m, n\}$, $n \leq k < m$, and $k \geq m$ separately. First we find an upper bound on secret DMT using (9).

For the first case $k < \min\{m, n\}$, $p > 0$ by Lemma 1 and $\mathbf{H}_E$ is not full column rank; i.e. $k < m$, then the secrecy capacity at high SNR is given by [14]

$$\begin{aligned} \tilde{C}_s(\mathrm{SNR}) &= \sum_{j: \sigma_j \geq 1} \log \sigma_j^2 + \log \left| \mathbf{I}_n + \frac{m\mathrm{SNR}}{p} \mathbf{H}_D \mathbf{H}_{\bar{E}}^\perp \mathbf{H}_D^\dagger \right| \\ &\quad + o(1), \end{aligned} \quad (13)$$

where $o(1) \to 0$ when $\mathrm{SNR} \to \infty$, $\mathbf{H}_{\bar{E}}^\perp \in \mathbb{C}^{m \times m}$ is the projection matrix onto $\text{Null}(\mathbf{H}_E)$, and $\sigma_j, j = 1, ..., \min\{m, n\} - p$, are the generalized singular values of matrices $\mathbf{H}_D$ and $\mathbf{H}_E$. To find the secret DMT we investigate the secrecy rate outage probability

$$P(\text{secrecy rate outage}) \quad (14)$$

$$\begin{aligned} &= P\left( \sum_{j: \sigma_j \geq 1} \log \sigma_j^2 + \log \left| \mathbf{I}_n + \frac{m\mathrm{SNR}}{p} \mathbf{H}_D \mathbf{H}_{\bar{E}}^\perp \mathbf{H}_D^\dagger \right| \right. \\ &\quad \left. + o(1) < r_s \log \mathrm{SNR} \right) \end{aligned} \quad (15)$$

$$\doteq P\left(\log\left|\mathbf{I}_n + \frac{m\mathrm{SNR}}{p}\mathbf{H}_D\mathbf{H}_E^\perp\mathbf{H}_D^\dagger\right| < r_s\log\mathrm{SNR}\right) \quad (16)$$

$$= \int...\int P\left(\log\left|\mathbf{I}_n + \frac{m\mathrm{SNR}}{p}\mathbf{H}_D\mathbf{H}_E^\perp\mathbf{H}_D^\dagger\right| < r_s\log\mathrm{SNR}\right.$$
$$\left.|\mathbf{H}_E^{(11)} = H_E^{(11)}, ..., \mathbf{H}_E^{(km)} = H_E^{(km)}\right)$$
$$\cdot \prod_{i=1}^{k}\prod_{j=1}^{m} f_{\mathbf{H}_E^{(ij)}}(H_E^{(ij)})dH_E^{(ij)} \quad (17)$$

For a fixed $\mathbf{H}_E = \mathbf{H}_E$, i.e. when all $\mathbf{H}_E^{(ij)} = H_E^{(ij)}$, $i = 1, ..., k, j = 1, ..., m$, the projection matrix $\mathbf{H}_E^\perp$ can be written as $\mathbf{H}_E^\perp = \mathbf{AA}^\dagger$. The matrix $\mathbf{A}$ is of size $m \times (m-k)$. We can write $\mathbf{A} = [a_1, ..., a_{m-k}]$, where the length-$m$ column vectors $a_j$ form an orthonormal basis for $\mathrm{Null}(\mathbf{H}_E)$. Let $\mathbf{H}_D = [\mathbf{r}_1^\dagger, ..., \mathbf{r}_n^\dagger]^\dagger$ be written in terms of length-$m$ row vectors $\mathbf{r}_i$, $i = 1, ..., n$. Then each entry of $(\mathbf{H}_D\mathbf{A})^{(ij)} = \langle \mathbf{r}_i, a_j \rangle$, $i = 1, ..., n, j = 1, ..., (m-k)$. The mean value of each entry is equal to $E\{\langle \mathbf{r}_i, a_j \rangle\} = 0$. We observe that the covariance $E\{\langle a_j^\dagger, \mathbf{r}_i^\dagger \rangle \langle \mathbf{r}_s, a_t \rangle\} = a_j^\dagger E\{\mathbf{r}_i^\dagger \mathbf{r}_s\}a_t$. The value $E\{\mathbf{r}_i^\dagger \mathbf{r}_s\} = 1$, if $i = s$, and it is zero if $i \neq s$. In addition to these, as the vectors are orthonormal, $a_j^\dagger E\{\mathbf{r}_i^\dagger \mathbf{r}_s\}a_t = 0$, if $j \neq t$ for any $i$ and $s$. Therefore, if $i = s$ and $j = t$, then $E\{\langle a_j^\dagger, \mathbf{r}_i^\dagger \rangle \langle \mathbf{r}_s, a_t \rangle\} = 1$; otherwise, it is equal to zero. Thus, $\mathbf{H}_D\mathbf{A}$ is a matrix, whose entries are i.i.d. Gaussian with zero mean and unit variance. Then we can write the probability in (17) as

$$P\left(\log\left|\mathbf{I}_n + \frac{m\mathrm{SNR}}{p}\mathbf{H}_D\mathbf{H}_E^\perp\mathbf{H}_D^\dagger\right| < r_s\log\mathrm{SNR}\right.$$
$$\left.|\mathbf{H}_E^{(11)} = H_E^{(11)}, ..., \mathbf{H}_E^{(km)} = H_E^{(km)}\right)$$
$$= P\left(\log\left|\mathbf{I}_n + \frac{m\mathrm{SNR}}{p}\mathbf{H}_D\mathbf{AA}^\dagger\mathbf{H}_D^\dagger\right| < r_s\log\mathrm{SNR}\right)$$
$$\doteq \mathrm{SNR}^{-d_{(m-k),n}(r_s)}.$$

In other words, this system is equivalent to an $(m-k) \times n$ MIMO with a well known DMT $d_{(m-k),n}(r_s)$ [28]. Substituting this value in (17), we observe that

$$P(\text{secrecy rate outage})$$
$$\doteq \int...\int \frac{1}{\mathrm{SNR}^{d_{(m-k),n}(r_s)}} \prod_{i=1}^{k}\prod_{j=1}^{m} f_{\mathbf{H}_E^{(ij)}}(H_E^{(ij)})dH_E^{(ij)}$$
$$= \mathrm{SNR}^{-d_{(m-k),n}(r_s)}$$

or $d_s(r_s) \le d_{(m-k),n}(r_s)$.

To attain secrecy we assume the source uses i.i.d. complex Gaussian codewords with covariance matrix $Q$ and transmits at rate $R^{(T)} = R_s^{(T)} + R^{(d)}$ bits/channel use, where $Q$ is the covariance matrix that attains the maximum in (12). Here the target secret communication rate is $R_s^{(T)}$ bits/channel use. Unlike the no CSIT case, the number of dummy codewords used for each secret message, $B = 2^{NR^{(d)}}$ is variable and equal to $2^{N\log|\mathbf{I}_k + \mathbf{H}_E Q\mathbf{H}_E^\dagger|}$. Then the main channel outage probability is equal to

$$P(\text{main channel outage})$$
$$= P\left(\log\left|\mathbf{I}_n + \mathbf{H}_D Q\mathbf{H}_D^\dagger\right| < R^{(T)}\right)$$

$$= P\left(\log\frac{\left|\mathbf{I}_n + \mathbf{H}_D Q\mathbf{H}_D^\dagger\right|}{\left|\mathbf{I}_k + \mathbf{H}_E Q\mathbf{H}_E^\dagger\right|} < r_s\log\mathrm{SNR}\right)$$
$$= P(\text{secrecy rate outage})$$
$$\doteq \mathrm{SNR}^{-d_{m-k,n}(r_s)}.$$

On the other hand, the probability of secrecy not achieved is

$$P(\text{secrecy not achieved})$$
$$\doteq P(I(\mathbf{X}; \mathbf{Y}_E) > R^{(d)}(\mathrm{SNR}))$$
$$= P(\log\left|\mathbf{I}_k + \mathbf{H}_E Q\mathbf{H}_E^\dagger\right| < I(\mathbf{X}; \mathbf{Y}_E)) \doteq 0.$$

Combining the upper and lower bounds (5) and (9) we conclude that the secret DMT is equal to $d_{m-k,n}(r_s)$ if $k < m$.

Note that it is possible to obtain the same DMT by setting $R^{(T)}(\mathrm{SNR}) = \log\left|\mathbf{I}_n + \mathbf{H}_D Q\mathbf{H}_D^\dagger\right|$, and the dummy information rate $R^{(d)}(\mathrm{SNR}) = \left[\log\left|\mathbf{I}_n + \mathbf{H}_D Q\mathbf{H}_D^\dagger\right| - R_s^{(T)}(\mathrm{SNR})\right]^+$. This guarantees a desired constant secrecy rate $R_s^{(T)}(\mathrm{SNR})$ and ensures that the main channel is never in outage. However, unlike the previous scheme, $P(\text{secrecy not achieved})$ is not on the order of 0.

For the second case $n \le k < m$, $p = 0$ by Lemma 1 and the high SNR secrecy capacity expression of [14] cannot be used directly. However, the converse and achievability in [14] can be extended to cover for $p = 0$, by deleting certain rows and columns in the generalized singular value decomposition [30]. Then the same secrecy capacity expression as in (13) holds with $p$ replaced by $p' = \min\{m-k, n\}$. We can follow the same steps in the previous case to calculate $P(\text{secrecy not achieved})$, $P(\text{main channel outage})$, and $P(\text{secrecy rate outage})$, and find the secret DMT to be $d_{(m-k),n}(r_s)$ for $n \le k < m$.

Finally for the last case, $k \ge m$, the secrecy capacity at high SNR is given by [14]

$$\lim_{\mathrm{SNR}\to\infty} \tilde{C}_s(\mathrm{SNR}) = \sum_{j:\sigma_j \ge 1}\log\sigma_j^2. \quad (18)$$

As the capacity expression does not grow with increasing SNR, it is easy to see that the secret DMT is a single point $(0,0)$. ∎

Note that, in the proof of Theorem 2 in the first achievable scheme, the number of dummy codewords, $B = 2^{N\log|\mathbf{I}_k + \mathbf{H}_E Q\mathbf{H}_E^\dagger|}$, is adapted to the source-eavesdropper channel. Thus, secrecy is always attained. Available CSIT improves the secret DMT with respect to secret DMT without CSIT and can also be used to guarantee no information is leaked to the eavesdropper when the destination receives the information correctly. Hence, main channel outage and secrecy rate outage events are the same. This is unlike the second proposed strategy where the main channel is never in outage, but information may be leaked to the eavesdropper or desired secrecy rate may not be attained.

In Fig. 1 secret DMT with CSIT is shown for $m = 3$, $n = 4$ and $k = 2$ in comparison to the secret DMT without CSIT and the DMT without secrecy constraints. The DMT without secrecy constraints, the secret DMT with CSIT and the secret DMT without CSIT are shown to be respectively equal to $d_{3,4}(r_s)$, $d_{1,4}(r_s)$, and $d_{1,2}(r_s)$. In this example secrecy constraints impose both multiplexing gain and diversity gain losses whether CSIT exists or not.

On the other hand, if CSIT is available, secrecy constraints do not always result in multiplexing gain loss with respect to the DMT without secrecy constraints. This is illustrated in Fig. 2 for which the source, the destination and the eavesdropper respectively have 4, 2 and 1 antennas each. In this case, the secret DMT with full CSIT is equal to $d_{3,2}(r_s)$, $r_s \in [0, 2]$, whereas the secret DMT with no CSIT is equal to $d_{3,1}(r_s)$, $r_s \in [0, 1]$. Note that the secret DMT with no CSIT always experiences a degrees of freedom loss, whereas secret DMT with full CSIT only experiences secret diversity gain loss but not secret multiplexing gain loss, if $m - k \geq n$.

In Fig. 3 we compare secrecy outage probability for a 2-antenna source, a 2-antenna destination, and a single antenna eavesdropper, $m = 2$, $n = 2$ and $k = 1$ using the secrecy capacity achieving scheme in (12). The secret multiplexing gain is assumed to be equal to 0.75; thus the secret diversity levels are equal to 0.25, if there is no CSIT, and 0.5 if CSIT is available. Fig. 3 confirms these results, from which we can observe the secret diversity to be approximately equal to the predicted values.

### B. The Zero-forcing Scheme

In this section we propose a simple *zero-forcing* method that achieves the full CSIT secret DMT as an alternative to the capacity achieving strategy studied in Theorem 2.

As $k \geq m$ results in a trivial secret DMT, we assume $k < m$; i.e. $\mathbf{H}_E$ is not full column rank. In the zero-forcing protocol we transmit the secret information in $\mathbf{U}$, which is a length-$(m-k)$ column vector, and send $\mathbf{X} = \mathbf{AU}$ at the transmitter, where $\mathbf{H}_E^{\perp} = \mathbf{A}\mathbf{A}^\dagger$. The received signals at the destination and the eavesdropper respectively become

$$\begin{aligned} \mathbf{Y}_D &= \mathbf{H}_D\mathbf{AU} + \mathbf{Z}_D, \\ \mathbf{Y}_E &= \mathbf{H}_E\mathbf{AU} + \mathbf{Z}_E = \mathbf{Z}_E. \end{aligned}$$

Stated differently, the destination observes an equivalent channel of $\mathbf{H}_D\mathbf{A}$, whereas the eavesdropper only observes noise because the secret message is transmitted in its null space. In this scheme, we only send secret messages at a fixed transmission rate $R_s^{(T)}(\mathrm{SNR})$ in the null space of the eavesdropper and the dummy information rate is set to zero.

As the receiver knows the transmit strategy, it is also informed about $\mathbf{A}$ and thus about the equivalent channel. Then for every realization of $\mathbf{A}$, the equivalent channel gain matrix still has i.i.d. complex Gaussian entries with zero mean and unit variance. Assuming the covariance matrix of $\mathbf{U}$ is $m\mathrm{SNR}\mathbf{I}_{m-k}/(m-k)$, the achievable secrecy rate (12) becomes

$$I(\mathbf{X};\mathbf{Y}_D) = I(\mathbf{U};\mathbf{Y}_D) = \log\left|\mathbf{I}_n + \mathbf{H}_D\mathbf{H}_E^{\perp}\mathbf{H}_D^\dagger\frac{m}{m-k}\mathrm{SNR}\right|$$

as $I(\mathbf{X};\mathbf{Y}_E) = 0$. In other words, secrecy is always attained, and the probability (7) is always zero. On the other hand, the main channel outage probability is equal to $P(I(\mathbf{X};\mathbf{Y}_D) <$

$R_s^{(T)}(\mathrm{SNR})$), for which the DMT can easily be shown to be $d_{m-k,n}(r_s)$ as in (14)-(17). This extends the results in [24] and [25] to secret DMT, which prove that the zero-forcing method is optimal in terms of secure degrees of freedom.

Note that for MIMO channels the source node can do beamforming in the direction of the destination, if CSI is available at the transmitter. Whether a secrecy constraint exists or not, beamforming in the direction of the destination only adds power gain to the achievable mutual information $I(\mathbf{X};\mathbf{Y}_D)$ or $\log\left|\mathbf{I}_n + \mathbf{H}_D Q\mathbf{H}_D^\dagger\right|$ term in (12) and does not change the DMT [28] or the secret DMT. However, when there are secrecy constraints, the transmitter CSI can be used to control the *beam direction* of the transmitted message. With this information, when the message is transmitted in the null space of the eavesdropper, the secret DMT changes significantly as illustrated in the zero-forcing protocol. In the zero-forcing scheme, as the secret messages are transmitted in the null space of $\mathbf{H}_E$, secrecy is always achieved.

### C. Artificial Noise

In [31] the authors suggest an artificial noise scheme to increase achievable secrecy rates. In the artificial noise scheme the source node sends its messages in the range space of $\mathbf{H}_D$ and sends extra noise in $\mathbf{H}_D$'s null space. Let $\mathbf{T}$ be an $m \times (m-n)$ matrix, whose columns form an orthonormal basis for Null($\mathbf{H}_D$), $\mathbf{V}$ is a length-$(m-n)$ column vector with i.i.d. complex Gaussian entries with zero mean, and $\mathbf{S}$ be a length-$m$ column vector that carries source messages. Then source sends

$$\mathbf{X} = \mathbf{S} + \mathbf{TV}.$$

As $\mathbf{V}$ is received in the null space of $\mathbf{H}_D$, the destination is not affected from this extra noise $\mathbf{V}$, but the eavesdropper is. Then the received signals at the destination and the eavesdropper are

$$\begin{aligned} \mathbf{Y}_D &= \mathbf{H}_D\mathbf{S} + \mathbf{Z}_D \\ \mathbf{Y}_E &= \mathbf{H}_E\mathbf{S} + \mathbf{H}_E\mathbf{TV} + \mathbf{Z}_E. \end{aligned}$$

We assume the vectors $\mathbf{S}$ and $\mathbf{V}$ are independent and respectively have the covariance matrices $\mathcal{E}\{\mathbf{SS}^\dagger\} = \mathrm{SNR}\mathbf{I}_m/2$ and $\mathcal{E}\{\mathbf{VV}^\dagger\} = m\mathrm{SNR}\mathbf{I}_{m-n}/[2(m-n)]$. Note that this choice satisfies the total power constraint as:

$$\begin{aligned} \mathrm{Tr}(\mathcal{E}\{\mathbf{X}^\dagger\mathbf{X}\}) &= \mathrm{Tr}(\mathcal{E}\{\mathbf{S}^\dagger\mathbf{S}\}) + \mathrm{Tr}(\mathcal{E}\{\mathbf{V}^\dagger\mathbf{T}^\dagger\mathbf{TV}\}) \\ &= \mathrm{Tr}(\mathcal{E}\{\mathbf{S}^\dagger\mathbf{S}\}) + \mathrm{Tr}(\mathcal{E}\{\mathbf{V}^\dagger\mathbf{V}\}) \\ &= \mathrm{Tr}(\mathcal{E}\{\mathbf{SS}^\dagger\}) + \mathrm{Tr}(\mathcal{E}\{\mathbf{VV}^\dagger\}) \\ &= m\mathrm{SNR}/2 + m\mathrm{SNR}/2. \end{aligned}$$

Then the achievable secrecy (3) becomes

$$R_s = \log\left|\mathbf{I}_n + \frac{\mathrm{SNR}}{2}\mathbf{H}_D\mathbf{H}_D^\dagger\right| - \log\frac{\left|\mathbf{K} + \frac{\mathrm{SNR}}{2}\mathbf{H}_E\mathbf{H}_E^\dagger\right|}{|\mathbf{K}|},$$

where $\mathbf{K} = \mathbf{I}_k + \frac{\mathrm{SNR}}{2(m-n)}\mathbf{H}_E\mathbf{TT}^\dagger\mathbf{H}_E^\dagger$.

Simulations suggest that the artificial noise scheme also achieves the secret DMT $d_{(m-k),n}(r_s)$ for $m = 2$, $n = k = 1$. A comparison between zero-forcing and artificial noise protocols is shown in Fig. 4 when the source has 2 antennas and the destination and the eavesdropper respectively have a single antenna each. The figure confirms that both schemes achieve a secret diversity 0.25, when the secret multiplexing gain is 0.75. The artificial noise scheme only necessitates the main channel CSIT to determine the codebook structure, whereas to do zero-forcing the instantaneous channel gain matrix of the eavesdropper is also required. However, the latter has a superior outage probability performance with respect to artificial noise. In zero-forcing the source concentrates its power in the null space of $\mathbf{H}_E$ and it is guaranteed that the eavesdropper does not get any information. Thus, an advantage of zero forcing is that the source does not have to employ a secret codebook or send dummy information.

## V. Conclusion

In this paper we study the MIMO wire-tap channel when there are stringent delay constraints and short-term power constraint. We define and find the *secret* DMT for arbitrary number of antennas at the source, the destination and the eavesdropper. First, we study no CSIT case with isotropic Gaussian codebook. Our results show that the eavesdropper decreases the degrees of freedom in the direct link, $\min\{m, n\}$, by the degrees of freedom in the source-eavesdropper channel, $\min\{m, k\}$. The secret DMT depends on the remaining degrees of freedom. Therefore, if $k \geq m$, then no degrees of freedom is left for secure communication. Otherwise, the secret DMT is equivalent to that of a $(m - k) \times (n - k)$ MIMO without secrecy constraints. Then we study the effect of transmitter CSI on secret DMT. We observe that unlike the DMT without secrecy constraints, the transmitter CSI changes the secret DMT and it becomes equivalent to the DMT of a $(m - k) \times n$ MIMO if $k < m$; otherwise no tradeoff exists between secret multiplexing and secret diversity. We also suggest a zero-forcing scheme, which achieves the secret DMT bound when CSIT is available, and compare it to the artificial noise scheme.

In this paper when there is CSIT, we assumed the source knows both the main channel CSI and the eavesdropper channel CSI to find the fundamental limits. When there is only main channel CSI the secure degrees of freedom is recently established in [27], but investigating the secret DMT remains to be an interesting open problem. Other possible future directions include finding the secret DMT for imperfect or partial CSI and finding an analytical expression for the artificial noise DMT.

## Appendix I
## Secrecy Rate Outage Probability, No CSIT

In Theorem 1 we need the probability of secrecy rate outage to calculate a lower bound on the probability of error. Thus, a lower bound on the probability of secrecy rate outage is sufficient. However, in this appendix we find both a lower bound and an upper bound on the secrecy rate outage probability to show that the bounds are tight. We first do the computations for $k < \min\{m, n\}$. We will discuss the case $k \geq \min\{m, n\}$ at the end of the proof. As for the no CSIT case of Section III we assume Gaussian codebooks and the input covariance matrix $Q = \text{SNR}\mathbf{I}_m$, it is sufficient that we find $P(\text{secrecy rate outage})$ under the same set of assumptions.

*Secrecy Rate Outage Lower Bound:* Define $\mathcal{E}_l = \{\mu_i > a, i = 1, ..., k\}$, where $a$ is a positive real constant and $\mathcal{E}_l^c$ denotes the complement of $\mathcal{E}_l$. Without loss of generality $a > 1$. Then we can write probability of secrecy rate outage as

$$
\begin{aligned}
&P(\text{secrecy rate outage}) \\
&= \quad P(\text{secrecy rate outage}|\mathcal{E}_l)P(\mathcal{E}_l) \\
&\quad + P(\text{secrecy rate outage}|\mathcal{E}_l^c)P(\mathcal{E}_l^c) \quad (19) \\
&\geq \quad P(\text{secrecy rate outage}|\mathcal{E}_l)P(\mathcal{E}_l). \quad (20)
\end{aligned}
$$

As $a$ is a constant, we find $P(\mathcal{E}_l)$ is also equal to a constant. At high SNR

$$
\begin{aligned}
&P(\text{secrecy rate outage}|\mathcal{E}_l) \\
&= \quad P\left(\frac{\prod_{i=1}^{L}(1 + \lambda_i \text{SNR})}{\prod_{i=1}^{k}(1 + \mu_i \text{SNR})} < \text{SNR}^{r_s} \Big| \mathcal{E}_l\right) \quad (21) \\
&\overset{(a)}{\geq} \quad P\left(\frac{\prod_{i=1}^{L}(1 + \lambda_i \text{SNR})}{(1 + a\text{SNR})^k} < \text{SNR}^{r_s}\right) \\
&\overset{(b)}{\geq} \quad P\left(\prod_{i=1}^{L}(1 + \lambda_i \text{SNR}) < \text{SNR}^{r_s + k}\right) \\
&\overset{(c)}{\doteq} \quad \text{SNR}^{-d_{m,n}(r_s + k)} \\
&\overset{(d)}{\doteq} \quad \text{SNR}^{-d_{m-k,n-k}(r_s)},
\end{aligned}
$$

where using the definition of $\mathcal{E}_l$ we substituted the minimum value for all $\mu_i$ in (21) to obtain $(a)$. $(b)$ follows because $(1 + a\text{SNR})^k \geq \text{SNR}^k$. Using the DMT results without secrecy constraints [28], $(c)$ and $(d)$ follow. As $P(\text{secrecy rate outage}) \doteq \text{SNR}^{-d_s(r_s)}$, we conclude that $d_s(r_s)$ is upper bounded by the DMT of an $(m-k) \times (n-k)$ MIMO system.

*Secrecy Rate Outage Upper Bound:* To argue that the secrecy rate outage lower bound is tight, we need a piecewise analysis, which depends on the secret multiplexing gain. We define $c_i$ as

$$c_i = -(m + n - 2k - 2i - 1)r_s + (m - k)(n - k) - i(i + 1),$$

for $i = 0, 1, ..., \min\{m, n\} - k - 1$. We also define the event $\mathcal{E}_{u,i} = \{\mu_k > c_i \log \text{SNR}\}$, and $\mathcal{E}_{u,i}^c$ as the complement of $\mathcal{E}_{u,i}$.

For any $i$, we can write

$$
\begin{aligned}
&P(\text{secrecy rate outage}) \\
&= \quad P(\text{secrecy rate outage}|\mathcal{E}_{u,i})P(\mathcal{E}_{u,i}) \\
&\quad + P(\text{secrecy rate outage}|\mathcal{E}_{u,i}^c)P(\mathcal{E}_{u,i}^c) \\
&\leq \quad P(\mathcal{E}_{u,i}) + P(\text{secrecy rate outage}|\mathcal{E}_{u,i}^c), \quad (22)
\end{aligned}
$$

where we have upper bounded both $P(\text{secrecy rate outage}|\mathcal{E}_{u,i})$ and $P(\mathcal{E}_{u,i}^c)$ with 1.

To calculate an upper bound on the first term in (22), we use an upper bound on the probability density function (pdf) of $\mu_k$. We obtain this bound using the joint pdf of the ordered eigenvalues $0 \leq \mu_1 \leq \mu_2 \leq ... \leq \mu_k$ of the matrix $\mathbf{H}_E \mathbf{H}_E^\dagger$

[28, Lemma 3], which is

$$p(\mu_1, ...\mu_k) = K_{m,k}^{-1} \prod_{i=1}^{k} \mu_i^{m-k} \prod_{i<j} (\mu_i - \mu_j)^2 e^{-\sum_{i=1}^{k} \mu_i},$$

where $K_{m,n}$ is a normalizing constant. Then,

$$
\begin{aligned}
p(\mu_k) &= \int_0^{\mu_k} ... \int_0^{\mu_2} p(\mu_1, ...\mu_k) d\mu_1 ... d\mu_{k-1} \\
&= K_{m,k}^{-1} \mu_k^{m-k} e^{-\mu_k} \\
&\quad \int_0^{\mu_k} ... \int_0^{\mu_2} \prod_{i=1}^{k-1} \mu_i^{m-k} \prod_{i<j} (\mu_i - \mu_j)^2 e^{-\sum_{i=1}^{k-1} \mu_i} \\
&\qquad d\mu_1 ... d\mu_{k-1} \\
&\overset{(e)}{\leq} K_{m,k}^{-1} \mu_k^{m-k} \mu_k^{k(k-1)} e^{-\mu_k} \\
&\quad \int_0^{\mu_k} ... \int_0^{\mu_2} \prod_{i=1}^{k-1} \mu_i^{m-k} e^{-\sum_{i=1}^{k-1} \mu_i} d\mu_1 ... d\mu_{k-1} \\
&\hspace{8cm} (23) \\
&\overset{(f)}{\leq} K_{m,k}^{-1} \mu_k^{m-k} \mu_k^{k(k-1)} e^{-\mu_k} [(m-k)!]^{k-1} \quad (24)
\end{aligned}
$$

where $(e)$ is because each $(\mu_i - \mu_j)^2 \leq \mu_k^2$, and there are $k(k-1)/2$ many $(\mu_i - \mu_j)^2$ terms involved. Before we write $(f)$ we first bound the innermost integral in (23) as

$$
\begin{aligned}
\int_0^{\mu_2} \mu_1^{m-k} e^{-\mu_1} d\mu_1 &= \gamma(m-k+1, \mu_2) \\
&\overset{(g)}{=} (m-k)! \left[ 1 - e^{-\mu_2} \left( \sum_{l=0}^{m-k} \frac{\mu_2^l}{l!} \right) \right] \\
&\leq (m-k)!,
\end{aligned}
$$

where $\gamma(.,.)$ is the lower incomplete gamma function. Note that for $(g)$ we used the series expansion of this function [32]. Applying this result repeatedly to all the integrals in (23) leads to $(f)$. Using this upper bound on the pdf of the largest eigenvalue $\mu_k$, we can now find an upper bound on $P(\mathcal{E}_{u,i})$. Let $C_i = c_i \log \mathrm{SNR}$ for short hand notation. Then,

$$
\begin{aligned}
&P(\mathcal{E}_{u,i}) \\
&= P(\mu_k > C_i) \\
&= \int_{C_i}^{\infty} p(\mu_k) d\mu_k \\
&\overset{(h)}{\leq} K_{m,k}^{-1} [(m-k)!]^{k-1} \int_{C_i}^{\infty} \mu_k^{m-2k+k^2} e^{-\mu_k} d\mu_k \\
&= K_{m,k}^{-1} [(m-k)!]^{k-1} \Gamma(m-2k+k^2+1, C_i) \quad (25) \\
&\overset{(i)}{=} K_{m,k}^{-1} [(m-k)!]^{k-1} e^{-C_i} \sum_{l=0}^{m-2k+k^2} \frac{C_i^l}{l!}, \quad (26)
\end{aligned}
$$

where we used (24) to obtain $(h)$. In (25) $\Gamma(.,.)$ denotes the upper incomplete Gamma function, and we used the series expansion of this function to obtain $(i)$ [32]. Then, it is easy to show that

$$P(\mathcal{E}_{u,i}) \overset{\cdot}{\leq} \mathrm{SNR}^{-c_i}, \quad (27)$$

as the $e^{-C_i}$ term in (26) determines the high SNR behavior of (26).

For the second term in (22) we show that

$$
\begin{aligned}
&P(\text{secrecy rate outage} | \mathcal{E}_{u,i}^c) \\
&= P \left( \frac{\prod_{i=1}^{L} (1 + \lambda_i \mathrm{SNR})}{\prod_{i=1}^{k} (1 + \mu_i \mathrm{SNR})} < \mathrm{SNR}^{r_s} | \mathcal{E}_{u,i}^c \right) \\
&\overset{(j)}{\leq} P \left( \frac{\prod_{i=1}^{L} (1 + \lambda_i \mathrm{SNR})}{(1 + (c_i \log \mathrm{SNR}) \mathrm{SNR})^k} < \mathrm{SNR}^{r_s} \right) \\
&\overset{(k)}{\leq} P \left( \prod_{i=1}^{L} (1 + \lambda_i \mathrm{SNR}) \right. \\
&\qquad \left. < (2 \max\{1, c_i\})^k (\log \mathrm{SNR})^k \mathrm{SNR}^{r_s+k} \right) \\
&\overset{(l)}{\leq} P \left( \prod_{i=1}^{L} (1 + \lambda_i \mathrm{SNR}) < A^k (\log \mathrm{SNR})^k \mathrm{SNR}^{r_s+k} \right) \\
&\overset{(m)}{=} \mathrm{SNR}^{-d_{m,n}(r_s+k)} \\
&\overset{\cdot}{=} \mathrm{SNR}^{-d_{(m-k),(n-k)}(r_s)}. \quad (28)
\end{aligned}
$$

In the above inequalities, $(j)$ is because the largest eigenvalue $\mu_k$ and hence all $\mu_i$'s are upper bounded by $c_i \log \mathrm{SNR}$ given $\mathcal{E}_{u,i}^c$. $(k)$ is due to the fact that

$$1 + (c_i \log \mathrm{SNR}) \mathrm{SNR} \leq 2 \max\{1, c_i\} (\log \mathrm{SNR}) \mathrm{SNR},$$

and $(l)$ follows because $c_i \leq (m-k)(n-k)$, for all $i = 1, ..., \min\{m, n\} - k - 1$, and we define $A = 2 \max\{1, (m-k)(n-k)\}$. Finally, $(m)$ is because $A^k (\log \mathrm{SNR})^k \mathrm{SNR}^{r_s+k}$ has the same multiplexing gain as $\mathrm{SNR}^{r_s+k}$, and thus the results in [28] apply.

Overall, substituting (27) and (28) into (22), using the definition of $c_i$, and

$$P(\text{secrecy rate outage}) \overset{\cdot}{\leq} \mathrm{SNR}^{-d_{(m-k),(n-k)}(r_s)}.$$

We can observe that this upper bound on probability of secrecy rate outage is the same as the lower bound we calculated above. We conclude that $P(\text{secrecy rate outage}) \overset{\cdot}{=} \mathrm{SNR}^{-d_s(r_s)} = \mathrm{SNR}^{-d_{(m-k),(n-k)}(r_s)}$ and the secret multiplexing gain satisfies $r_s \leq \min\{m, n\} - k$ for $k < \min\{m, n\}$.

If $k \geq \min\{m, n\}$, then $P(\text{secrecy rate outage} | \mathcal{E}_l)$ in (20) takes a constant value and does not decay with SNR. As $P(\mathcal{E}_l)$ is also equal to a constant, $P(\text{secrecy rate outage})$ is lower bounded by a fixed value in $(0, 1]$. Thus, we conclude that when $k \geq \min\{m, n\}$, the secret DMT reduces to the single point $(0, 0)$.

## REFERENCES

[1] A. D. Wyner, "The wire-tap channel," *The Bell System Technical Journal*, vol. 54, p. 1355, October 1975.

[2] I. Csiszár and J. Körner, "Broadcast channels with confidential messages," *IEEE Transactions on Information Theory*, vol. 24, no. 3, p. 339, May 1978.

[3] S. K. Leung-Yan-Cheong and M. E. Hellman, "The Gaussian wire-tap channel," *IEEE Transactions on Information Theory*, vol. 24, no. 4, p. 451, July 1978.

[4] P. K. Gopala, L. Lai, and H. E. Gamal, "On the secrecy of capacity of fading channels," *IEEE Transactions on Information Theory*, vol. 54, no. 10, p. 4687, October 2008.

[5] Y. Liang, H. V. Poor, and S. Shamai, "Secure communication over fading channels," *IEEE Transactions on Information Theory, Special Issue on Information Theoretic Security*, vol. 54, p. 2470, June 2008.

[6] E. Tekin, S. Serbetli, and A. Yener, "On secure signaling for the Gaussian multiple access wire-tap channel," october 2005, pp. 1747 – 1751.

[7] E. Tekin and A. Yener, "The general Gaussian multiple access and two-way wire-tap channels: Achievable rates and cooperative jamming," *IEEE Transactions on Information Theory, Special Issue on Information Theoretic Security*, vol. 54, no. 6, p. 2735, June 2008.

[8] R. Liu, I. Maric, P. Spasojevic, and R. D. Yates, "Discrete memoryless interference and broadcast channels with confidential messages: Secrecy rate regions," *IEEE Transactions on Information Theory*, p. 2493, June 2008.

[9] A. Khisti, A. Tchamkerten, and G. Wornell, "Secure broadcasting over fading channels," *IEEE Transactions on Information Theory*, vol. 54, no. 6, pp. 2453 –2469, june 2008.

[10] L. Lai and H. E. Gamal, "The relay-eavesdropper channel: Cooperation for secrecy," *IEEE Transactions on Information Theory*, vol. 54, p. 4005, September 2008.

[11] M. Yuksel and E. Erkip, "The relay channel with a wire-tapper," in *Proceedings of 41st Conference of Information Sciences and Systems*, March 2007.

[12] P. Parada and R. Blahut, "Secrecy capacity of SIMO and slow fading channels," in *Proceedings of IEEE International Symposium on Information Theory*, 2005.

[13] S. Shafiee, N. Liu, and S. Ulukus, "Towards the secrecy capacity of the Gaussian MIMO wire-tap channel: The 2-2-1 channel," *IEEE Transactions on Information Theory*, vol. 55, no. 9, pp. 4033 –4039, September 2009.

[14] A. Khisti and G. W. Wornell, "Secure transmission with multiple antennas- II: The MIMOME wiretap channel," to appear, IEEE Transactions on Information Theory.

[15] F. Oggier and B. Hassibi, "The secrecy capacity of the MIMO wiretap channel." [Online]. Available: http://arxiv.org/abs/0710.1920

[16] Z. Li, W. Trappe, and R. Yates, "Secret communication via multi-antenna transmission," in *Proceedings of 41st Conference of Information Sciences and Systems*, Baltimore, MD, March 2007.

[17] R. Liu and H. V. Poor, "Multiple antenna secure broadcast over wireless networks," in *Proceedings of the First International Workshop on Information Theory for Sensor Networks*, June 18 - 20 2007.

[18] R. Bustin, R. Liu, H. V. Poor, and S. S. (Shitz), "An MMSE approach to the secrecy capacity of the MIMO Gaussian wiretap channel," *EURASIP Journal on Wireless Communications and Networking*, vol. 2009, pp. 1–8, 2009.

[19] J. Barros and M. R. D. Rodrigues, "Secrecy capacity of wireless channels," in *Proceedings of IEEE International Symposium on Information Theory*, 2006.

[20] K. Khalil, O. O. Koyluoglu, H. E. Gamal, and M. Youssef, "Opportunistic secrecy with a strict delay constraint," July 2009, submitted to IEEE Transactions on Information Theory. [Online]. Available: http://arxiv.org/abs/0907.3341

[21] Y. Liang, G. Kramer, H. V. Poor, and S. Shamai, "Compound wiretap channels," *EURASIP Journal on Wireless Communications and Networking, Special Issue on Wireless Physical Layer Security*, vol. 2009, 2009.

[22] O. O. Koyluoglu, H. E. Gamal, L. Lai, and H. V. Poor, "Interference alignment for secrecy," to appear in IEEE Transactions on Information Theory.

[23] T. Gou and S. A. Jafar, "On the secure degrees of freedom of wireless X networks," in *Proceedings of 46th Annual Allerton Conference on Communication, Control, and Computing*, 2008.

[24] M. Kobayashi, M. Debbah, and S. S. (Shitz), "Secured communication over frequency-selective fading channels: A practical vandermonde precoding," *EURASIP Journal on Wireless Communications and Networking*, 2009.

[25] M. Kobayashi, Y. Liang, S. S. (Shitz), and M. Debbah, "On the compound MIMO broadcast channels with confidential messages," in *Proceedings of IEEE International Symposium on Information Theory*, 2009.

[26] X. He and A. Yener, "Providing secrecy with unstructured codes: tools and applications to two-user Gaussian channels," July 2009, submitted to IEEE Transactions on Information Theory. [Online]. Available: http://arxiv.org/abs/0907.5388

[27] ——, "MIMO wiretap channels with arbitrarily varying eavesdropper channel states," July 2010, submitted to IEEE Transactions on Information Theory. [Online]. Available: http://arxiv.org/abs/1007.4801

[28] L. Zheng and D. N. C. Tse, "Diversity and multiplexing: A fundamental tradeoff in multiple-antenna channels," *IEEE Transactions on Information Theory*, vol. 49, p. 1073, May 2003.

[29] A. Thangaraj, S. Dihidar, A. R. Calderbank, S. W. McLaughlin, and J. Merolla, "Applications of LDPC codes to the wiretap channel," *IEEE Transactions on Information Theory*, vol. 53, no. 8, pp. 2933–2945, August 2007.

[30] C. C. Paige and M. A. Saunders, "Towards a generalized singular value decomposition," *SIAM Journal on Numerical Analysis*, vol. 18, p. 398, June 1981.

[31] S. Goel and R. Negi, "Guaranteeing secrecy using artificial noise," *IEEE Transactions on Wireless Communications*, vol. 7, no. 6, p. 2180, June 2008.

[32] I. S. Gradshteyn and I. M. Ryzhik, *Table of integrals, series, and products*, 7th ed. Academic press, 2007.

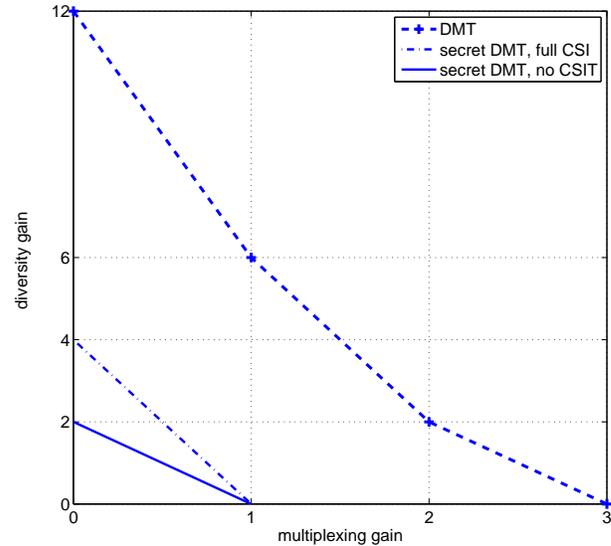

Fig. 1. The source, the destination and the eavesdropper respectively have 3, 4 and 2 antennas. The DMT with no secrecy constraints, secret DMT with transmitter and receiver CSI and, secret DMT with receiver CSI only are shown.

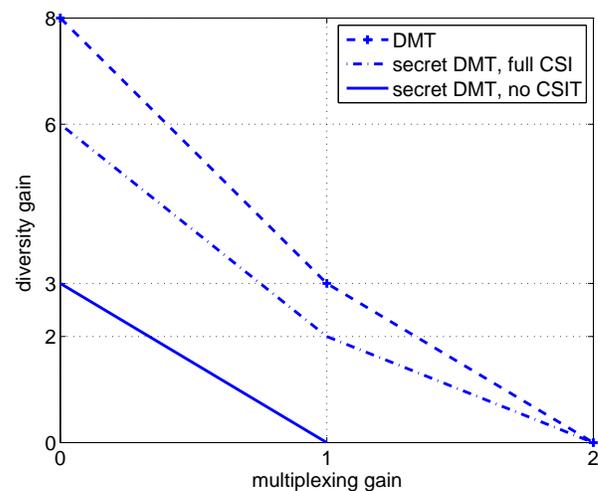

Fig. 2. The source, the destination and the eavesdropper respectively have 4, 2 and 1 antennas. The DMT with no secrecy constraints, secret DMT with transmitter and receiver CSI and, secret DMT with receiver CSI only are shown.

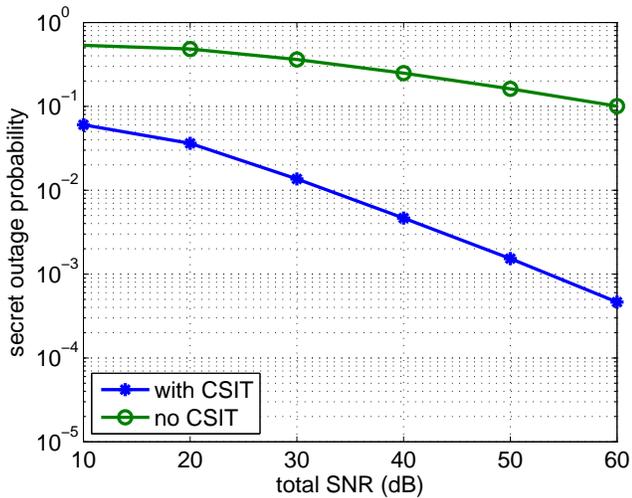

Fig. 3. The source, the destination and the eavesdropper respectively have 2, 2 and 1 antennas, $r_s = 0.75$.

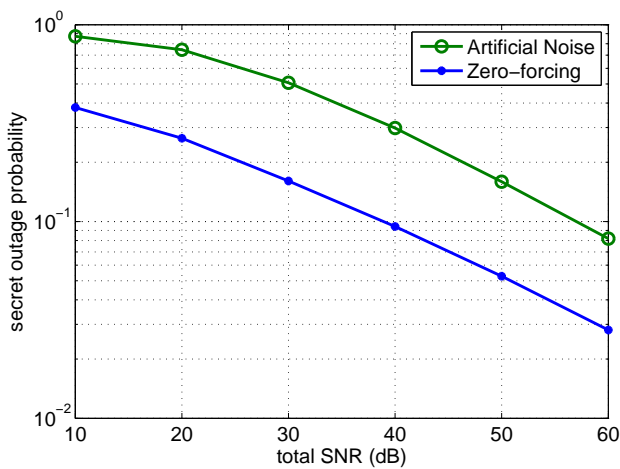

Fig. 4. The source has 2 antennas, and the destination and the eavesdropper each have a single antenna, $r_s = 0.75$.

PLACE PHOTO HERE

**Melda Yuksel** (S'98-M'07) received the Ph.D. degree in electrical engineering at Polytechnic University, Brooklyn, NY, in August 2007, and the B.S. degree in electrical and electronics engineering from Middle East Technical University, Ankara, Turkey, in 2001.

She joined TOBB University of Economics and Technology, Ankara, Turkey, in Fall 2007. In 2004, she was a summer researcher in Mathematical Sciences Research Center, Bell-Labs, Lucent Technologies, Murray Hill, NJ. Her research interests include communication theory and information theory and more specifically cooperative communications, network information theory, and information-theoretic security over communication channels.

Melda Yuksel is the recipient of the best paper award at the Communication Theory Symposium of the 2007 IEEE International Conference on Communications.

PLACE PHOTO HERE

**Elza Erkip** (S'93-M'96-SM'05-F'11) received the B.S. degree in electrical and electronics engineering from the Middle East Technical University, Ankara, Turkey, and the M.S. and Ph.D. degrees in electrical engineering from Stanford University, Stanford, CA. Currently, she is an associate professor of electrical and computer engineering at Polytechnic Institute of New York University, Brooklyn. In the past, she has held positions at Rice University, Houston, TX, and at Princeton University, Princeton, NJ. Her research interests are in information theory, communication theory, and wireless communications.

Dr. Erkip received the National Science Foundation CAREER Award in 2001, the IEEE Communications Society Rice Paper Prize in 2004, and the ICC Communication Theory Symposium Best Paper Award in 2007. She co-authored a paper that received the ISIT Student Paper Award in 2007. Currently, she is an associate editor of the IEEE TRANSACTIONS ON INFORMATION THEORY and a Technical Co-Chair of WiOpt 2011. She was an associate editor of the IEEE TRANSACTIONS ON COMMUNICATIONS during 2006-2009, a publications editor of the IEEE TRANSACTIONS ON INFORMATION THEORY during 2006-2009 and a guest editor of IEEE Signal Processing Magazine in 2007. She was the co-chair of the GLOBECOM Communication Theory Symposium in 2009, the publications chair of ITW Taormina in 2009, the MIMO Communications and Signal Processing Technical Area chair of the Asilomar Conference on Signals, Systems, and Computers in 2007, and the technical program co-chair of the Communication Theory Workshop in 2006.